\documentclass[lettersize,journal]{IEEEtran}
\usepackage{amsmath,amsfonts}
\usepackage{algorithmic}
\usepackage{algorithm}
\usepackage{array}
\usepackage[caption=false,font=normalsize,labelfont=sf,textfont=sf]{subfig}
\usepackage{textcomp}
\usepackage{stfloats}
\usepackage{booktabs}
\usepackage{multirow}
\usepackage{tabularx}
\usepackage{makecell}
\usepackage{url}
\usepackage{verbatim}
\usepackage{graphicx}
\usepackage{cite}
\begin{document}

\title{WISV: Wireless-Informed Semantic Verification for Distributed Speculative Decoding in Device–Edge LLM Inference}

\author{Zixuan Liu, Zhiyong Chen, \emph{Senior Member, IEEE}, Nan Xue, Shengkang Chen, Jiangchao Yao,\\ Meixia Tao, \emph{Fellow, IEEE}, and Wenjun Zhang, \emph{Fellow, IEEE}
\thanks{The authors are with the Cooperative Medianet Innovation Center and the Department of Electronic Engineering, Shanghai Jiao Tong University, Shanghai 200240, China. (e-mail:\{lzx20020405, zhiyongchen, nan.xue, Chen20, sunarker, mxtao, zhangwenjun\}@sjtu.edu.cn). \emph{(Corresponding author: Zhiyong Chen.)}}
}



\maketitle

\begin{abstract}
While distributed device–edge speculative decoding enhances resource utilization across heterogeneous nodes, its performance is often bottlenecked by conventional token-level verification strategies. Such rigid alignment leads to excessive rejections, significantly diminishing the accepted sequence length and increasing interaction rounds under fluctuating wireless conditions.
In this paper, we propose \textbf{WISV} (\textbf{W}ireless-\textbf{I}nformed \textbf{S}emantic \textbf{V}erification), a novel distributed speculative decoding framework that goes beyond strict token-level matching via a channel-aware semantic acceptance policy. WISV integrates a lightweight decision head into the edge-side target LLM to dynamically evaluate speculative tokens by synthesizing high-dimensional hidden representations with instantaneous channel state information (CSI). To optimize the trade-off between verification fidelity and communication overhead, we further design two tailored communication protocols: full-hidden upload and mismatch-first selective-hidden upload. Extensive simulations using a 1B drafter and an 8B target model demonstrate that WISV achieves up to a 60.8\% increase in accepted length, a 37.3\% reduction in interaction rounds, and a 31.4\% improvement in end-to-end latency compared to vanilla speculative decoding across tested settings, while maintaining a negligible task accuracy drop ($<$1\%). Finally, we validate WISV on a hardware testbed comprising an NVIDIA Jetson AGX Orin and an A40-equipped server, confirming its real-world efficacy in accelerating edge-deployed LLM inference.
\end{abstract} 

\section{Introduction}
\IEEEPARstart{R}{ecent} years have seen a rapid migration of large language model (LLM) inference from centralized clouds to distributed device–edge architectures, a shift necessitated by stringent latency requirements and the hardware limitations of standalone devices \cite{MEI4LLM,OnDeviceLMReview,OnDeviceAISurvey,EdgeEffLLMSurvey,AutonomousEdgeAI,EAI,Pushing,WDMoE, Petals}. While distributed execution alleviates device-side resource constraints and improves system scalability, its efficiency is fundamentally tethered to the underlying wireless networks. To further accelerate autoregressive decoding, recent advancements have introduced device–edge \emph{speculative decoding}, where a lightweight on-device drafter collaborates with a powerful edge-based verifier \cite{SD,SS,DSD,DSSD}. However, in wireless networks, the high-dimensional data exchange and unpredictable channel dynamics often lead to frequent verification failures and increased interaction rounds \cite{Neurosurgeon, Edgent, bottlenet++, NestDNN, EarlyExitSurvey, SplitNN}. These overheads not only exacerbate communication latency but also risk negating the acceleration gains of speculative execution, positioning the stochastic nature of wireless channels as a critical performance bottleneck \cite{MEI4LLM,JD,AJ,FLy}.

Motivated by this observation, the aim of this work is to understand how token acceptance in distributed speculative decoding should be designed under wireless communication constraints. Specifically, we first seek to characterize how the trade-off between semantic fidelity and link-dependent communication cost affects the acceptance decision, rather than relying on strict token-level matching. Second, we propose WISV, a wireless-informed semantic verification framework that augments the target model with a lightweight decision head to determine token acceptance by jointly leveraging semantic features and channel state information (CSI). 
\subsection{Related Work}
Speculative decoding has emerged as a promising approach to reduce LLM decoding latency. In this paradigm, a lightweight draft model proposes multiple candidate tokens, which are then verified in parallel by a larger target model, thereby amortizing the cost of autoregressive generation \cite{SD}. Related approaches include speculative sampling, which extends this idea to sampling-based decoding while preserving the target distribution \cite{SS}, and hierarchical or staged speculative decoding, which improves verification efficiency and is particularly suitable for small-batch or on-device inference \cite{overcoming,accelerating}. Other methods go beyond the classic draft-target design: Medusa predicts multiple future tokens in parallel \cite{medusa}, EAGLE performs feature-level speculative sampling \cite{eagle, eagle2, eagle3}, and self-speculative decoding eliminates the need for an external draft model by generating drafts via a faster internal path \cite{DraftVerify}. High-throughput LLM serving systems further enhance GPU utilization through efficient KV-cache management and scheduling for large-scale multi-user decoding \cite{vLLM,Orca,SarathiServe,DistServe,EfficientScaling,DeepSpeedInference,FlexGen}. Despite these advances, most decoding acceleration methods are primarily studied in colocated or server-centric settings and do not account for interactive communication overhead in wireless device–edge deployments \cite{MEI4LLM,DSD}.

To overcome the limitations of single-node speculative decoding, recent studies have explored distributed designs, where a lightweight drafter runs on a resource-constrained device and the target model resides on an edge or cloud server \cite{DSD,DSSD}. DuoDecoding further co-optimizes the draft budget with hardware characteristics \cite{duo}. Such distributed designs mitigate scalability bottlenecks of colocated decoding and better exploit on-device compute resources \cite{MEI4LLM,OnDeviceLMReview}. However, this separation introduces a new challenge: every verification now interacts with wireless transmission cost, and the optimal acceptance policy is no longer determined by token alignment alone \cite{MEI4LLM,DSD,DSSD}.

Existing distributed speculative decoding schemes often rely on strict distribution-matching verification to preserve the target model's output distribution \cite{SD,SS,DSD}. In wireless device–edge deployments, this strictness can be a liability. Draft tokens that are semantically acceptable may be rejected due to token-level mismatches, shortening accepted spans and triggering extra interaction rounds \cite{JD}. This is particularly problematic under bandwidth-limited or rapidly varying wireless channels, where each additional interaction incurs not only transmission overhead but also latency uncertainty \cite{MEI4LLM}. Recent lossy speculative decoding studies have shown that tolerating certain token-level mismatches can improve practical speedups and accepted length without degrading downstream response quality \cite{JD,AJ,FLy}. Yet, these methods are mostly designed for single-host acceleration or task-specific quality trade-offs and do not explicitly incorporate wireless communication cost into the acceptance policy \cite{MEI4LLM,DSD,DSSD}.

\subsection{Contributions}
This paper is motivated by a simple but consequential observation: in wireless device-edge LLM generation, token acceptance is not merely a binary ''correct/incorrect'' decision, but a trade-off under communication cost. Accepting a draft token reduces interaction rounds and transmission time, while rejecting it may improve token-level alignment at the cost of additional device–edge exchanges. Therefore, the optimal acceptance decision must consider both semantic fidelity and link-dependent communication penalties, naturally characterized by CSI and wireless dynamics \cite{MEI4LLM}.

Guided by this insight, we propose WISV in this paper, a wireless-informed semantic verification framework for distributed speculative decoding. WISV introduces a lightweight decision head attached to the target model, which performs token-level acceptance by jointly leveraging semantic consistency and wireless channel information. This design allows the system to selectively tolerate harmless mismatches while rejecting critical ones, thereby increasing the accepted token length per verification cycle and reducing interaction rounds.

To enable practical deployment under realistic uplink constraints, WISV is coupled with device–edge communication protocols that optimize the transmission of hidden states and token information. Together, these components form an end-to-end framework that explicitly balances semantic correctness with wireless communication cost, bridging distributed speculative decoding and judge-style verification in a unified device–edge setting.

The contributions of this paper are summarized as follows:
\begin{itemize}
\item \textbf{Wireless-Informed Semantic Verification Framework:} We propose WISV, which replaces rigid token-matching with a CSI-aware decision head. By coupling wireless link quality with semantic importance, WISV effectively mitigates the over-rejection problem in volatile wireless environments.

\item \textbf{Communication-Computation Co-optimized Protocols:} We develop two device–edge protocols: a full-hidden upload protocol for direct semantic judging, and a mismatch-first selective-hidden upload protocol that transmits only necessary hidden states after locating mismatches. These protocols substantially reduce uplink overhead while preserving decision quality, making WISV practical under realistic wireless constraints.
\item \textbf{Wireless-aware supervised data construction:} We introduce a training pipeline that captures token importance and incorporates channel-state information through a cost-aware relabeling procedure, enabling the decision head to learn acceptance policies that balance semantic correctness and communication cost.
\item \textbf{Hardware testbed validation:} We implement WISV on NVIDIA Jetson AGX Orin devices and an edge server, demonstrating that it consistently shortens end-to-end latency by up to 35.41\% compared with vanilla speculative decoding baselines.
\end{itemize}

The rest of the paper is organized as follows. Section II presents the principles of speculative decoding. The system model of WISV is introduced in Section~\ref{sec:system}. The problem formulation and performance metrics are introduced in Section ~\ref{sec:problem}. The realization of CSI-Aware semantic verification is presented in Section~\ref{sec:semantic}. Section~\ref{sec:sim} shows extensive simulation results. Finally, the hardware testbed experiments are presented in Section ~\ref{sec:hardware} and conclusions are drawn in Section~\ref{sec:conclusion}.

\section{Speculative Decoding}
\subsection{Speculative Decoding and Sampling}
Speculative decoding accelerates autoregressive generation by allowing a lightweight draft model to propose a block of $K$ tokens, which are then verified by a stronger target model in one forward pass per block \cite{SD}. 
Let $p_T(\cdot \mid x)$ and $p_D(\cdot \mid x)$ denote the token level distributions of the target and draft models given prefix $x$, respectively. 
In a common \emph{greedy verification} variant, a draft token is accepted if it matches the target model's argmax at the corresponding position; otherwise, the process stops at the first mismatch, and a new speculation round begins.

To preserve the target distribution beyond greedy decoding, \emph{speculative sampling} replaces deterministic matching with a modified rejection sampling rule \cite{SS}. 
Let $p_T(\cdot\mid x)$ and $p_D(\cdot\mid x)$ denote the next-token distributions of the target and drafter given prefix $x$. Given a draft token $y \sim p_D(\cdot\mid x)$, the verifier computes the acceptance probability
\begin{equation}
\alpha(x,y) = \min\left\{1,\ \frac{p_T(y\mid x)}{p_D(y\mid x)}\right\},
\label{eq:ss_accept_prob}
\end{equation}
and accepts $y$ with probability $\alpha(x,y)$; otherwise, it samples from a \emph{residual} distribution
\begin{equation}
r(\cdot\mid x) \ \propto\ p_T(\cdot\mid x) - \alpha(x,\cdot)\,p_D(\cdot\mid x),
\label{eq:ss_residual}
\end{equation}
which ensures that the emitted token is distributed exactly as $p_T(\cdot\mid x)$ (up to numerical issues and any practical approximation of $p_D$ transmission) \cite{SS}.

For a $K$-token speculative block, the procedure is applied sequentially. The drafter first proposes $(y_t,\ldots,y_{t+K-1})$, and the verifier evaluates $\alpha(x_t,y_t)$ at the first position. If accepted, the prefix is updated to $x_{t+1}$; otherwise, a token is sampled from the corresponding residual distribution. This process continues until the first rejection, after which the remaining tokens are discarded and a new speculation round begins \cite{SS}. Under this mechanism, the expected accepted length is determined by the per-position acceptance probabilities, which depend on the alignment between $p_D$ and $p_T$. As a result, decoding efficiency is closely tied to the accepted-length distribution and the resulting number of interaction rounds.

\begin{figure*}[t]
    \centering
    \includegraphics[width=1\linewidth]{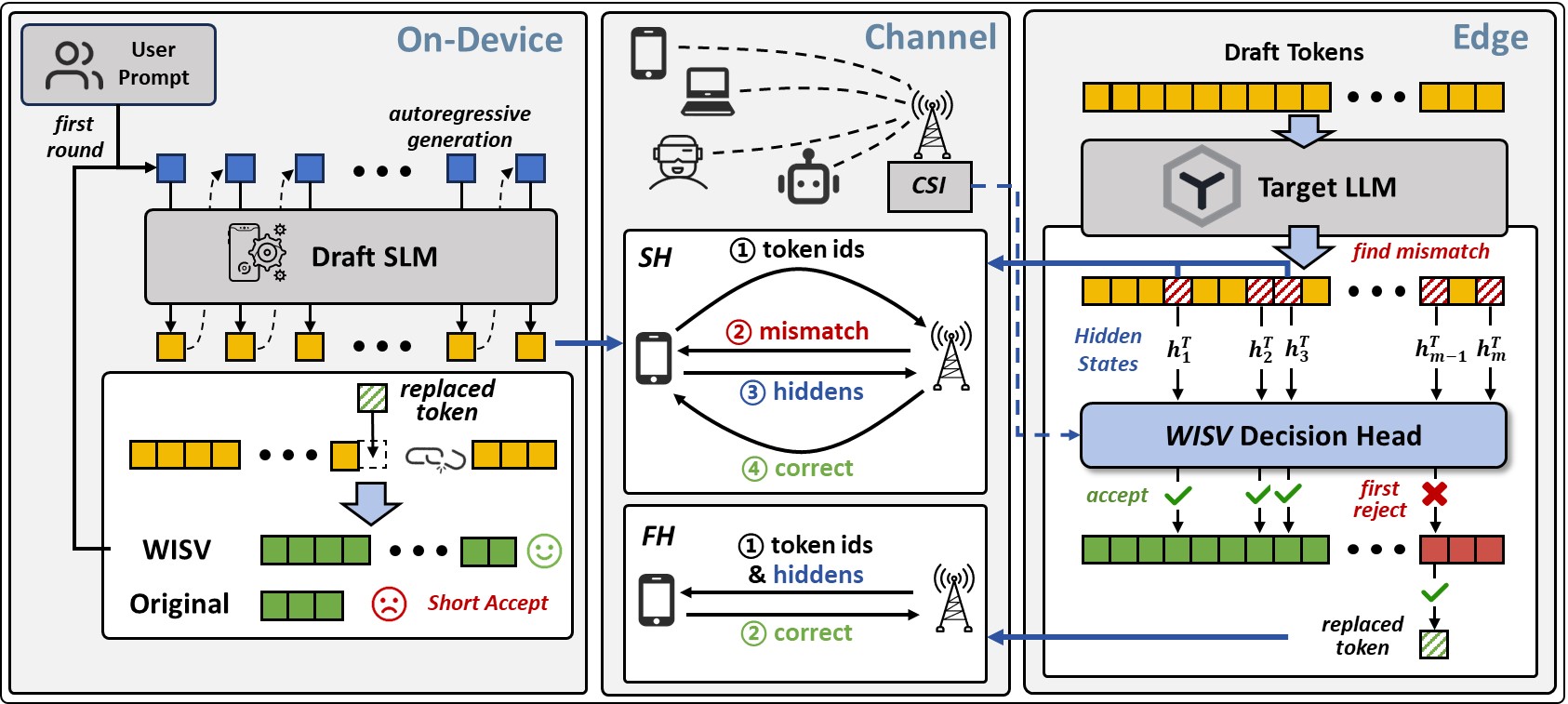}
    \caption{System model for the proposed WISV framework. }
    \label{fig:system}
\end{figure*}

\subsection{Distributed Speculative Decoding}
Recent works extend speculative decoding to \emph{distributed} device-edge or edge-cloud settings, where a lightweight drafter runs on the device while a stronger target model resides at the edge or cloud \cite{DSD,SLED,DSSD}. 
DSD introduces a distributed framework with adaptive speculation control to improve throughput under heterogeneous deployments \cite{DSD}. SLED focuses on multi-device edge serving by batching verification requests to enhance system capacity \cite{SLED}. DSSD further reduces communication overhead by splitting verification across device and edge to avoid repeated transmissions of large probability payloads \cite{DSSD}.

While these systems demonstrate the importance of protocol and system-level co-design, their verification mechanisms remain largely based on strict token-level alignment or distribution matching. As a result, semantically acceptable draft tokens may still be rejected, shortening accepted spans and increasing interaction rounds, a problem that becomes more pronounced under wireless communication constraints.

\subsection{Judge Decoding}
To relax strict token-level matching, recent works propose judge-style verification mechanisms that accept semantically valid tokens even when they do not exactly match the target model’s preferred output. Judge Decoding introduces a lightweight module to assess token validity based on target model representations, reducing unnecessary rejections and improving decoding efficiency \cite{JD}. AutoJudge further automates supervision construction and enables lossy acceptance with controlled quality trade-offs \cite{AJ}.

However, existing judge-based approaches are primarily designed for single-host or communication-agnostic settings. Their acceptance policies do not explicitly consider wireless link variability or the system cost of additional device–edge interactions, limiting their effectiveness in wireless distributed inference scenarios.

\section{System Architecture and Workflow of WISV}
\label{sec:system}
This section presents the proposed device-edge co-inference architecture and its operational workflow, incorporating token-block speculation, edge verification, and CSI-aware semantic acceptance. Subsequently, we present two communication protocols designed to balance uplink overhead against interaction latency.
\subsection{Architecture and Workflow}
\label{sec:arch_workflow}
As illustrated in Fig. 1, we consider a device–edge co-inference framework consisting of a user equipment (UE) and a base station (BS) equipped with an edge server. A lightweight drafter model is deployed on the UE, while a high-capacity target model resides on the edge server. The UE and the edge server interact over a wireless channel, with the instantaneous condition at interaction round $r$ characterized by the CSI, denoted as $\mathbf{c}^{r}$. The objective is to accelerate end-to-end autoregressive generation under wireless constraints by minimizing both the target-model's decoding workload and the total number of interaction rounds, all while preserving the output quality.

The operational workflow is detailed as follows:

1) \textbf{Token-block speculation:} At each round $r$, given a committed prefix $\mathbf{y}$ of length $t$, the drafter generates a block of $K$ candidate tokens
$\tilde{\mathbf{y}}_{t:t+K-1} = (\tilde{y}_t,\ldots,\tilde{y}_{t+K-1})$ via autoregressive decoding. Since the subsequent verification in WISV adopts a greedy verification strategy to localize mismatch tokens, the UE only transmits the token IDs and hidden states $\tilde{\mathbf{y}}_{t:t+K-1}$ to the edge server. This uplink transmission defines the speculation window for edge-side verification.

2) \textbf{Edge verification and mismatch localization:} Upon receiving the draft block, the edge performs a forward pass of the target model, conditioned on the prefix and the speculative draft tokens. This process yields decoding logits and hidden representations for each position. In greedy verification, mismatches are determined based on argmax tokens, pinpointing positions where semantic acceptance decisions could potentially extend the accepted sequence.

3) \textbf{Wireless-informed semantic acceptance via a decision head:} Conventional distribution-matching verification often rejects tokens that are semantically congruent but misaligned at the token level, which curtails the accepted span and increases interaction frequency. To address this, we integrate a lightweight decision head into the edge-side target model. For each flagged position $i \in \{t,\ldots,t+K-1\}$, the decision head takes three inputs:
$(i)$ the drafter hidden state $\mathbf{h}^{D}_i$, encoding the drafter's local semantic evidence;
$(ii)$ the target hidden state $\mathbf{h}^{T}_i$, representing the target model's contextual assessment; and
$(iii)$ the CSI feature vector $\mathbf{c}^{r}$, reflecting the current wireless link condition.

The head outputs a rejection probability for each draft token. By treating CSI as a first-class input, the system dynamically balances semantic risk against communication overhead, selectively accepting semantically valid tokens when the cost of re-transmission is high.

4) \textbf{Rollback-triggered feedback and iteration:} The edge determines the earliest rejected position $\hat{j}$ within the window. Tokens preceding $\hat{j}$ are accepted and committed. At position $\hat{j}$, the edge selects a corrected token $\,y_{\hat{j}}^{\star}\,$ from the target logits and returns a feedback message containing the index $\hat{j}$ and the corrected token ID $y_{\hat{j}}^{\star}$. The UE then updates its local prefix and resumes drafting from position $\hat{j}+1$. This cycle repeats until sequence is complete. If all tokens in the draft window are accepted, the edge provides the next predicted token to sustain the generation flow.

\subsection{Two mechanisms for obtaining drafter hidden states.} 
\label{sec:SHFH} 
A practical challenge is enabling the edge to acquire the drafter hidden states $\mathbf{h}^{D}_i$ with minimal uplink overhead. We propose two protocol options: 

\emph{(1) Full-hidden upload:} The UE transmits token IDs together with \emph{all} drafter hidden states for the $K$ draft tokens in the speculation window. This enables immediate evaluation by the decision head without additional interactions, but incurs an uplink payload that scales with the window size $K$ and hidden dimension.

\emph{(2) Selective-hidden upload:} The UE first transmits only token IDs. After mismatch localization at the edge, the edge requests hidden states only for selected positions, and the UE uploads them on demand. This reduces uplink traffic when only a small subset of positions is queried, at the cost of an additional interaction round.

These two mechanisms provide a flexible trade-off between uplink payload and interaction latency, and can be adapted to different CSI regimes and system constraints. The overall WISV decoding procedure with dual hidden-state protocols is summarized in Algorithm~\ref{alg:WISV}.

Compared with strict distribution-matching speculative decoding, the proposed CSI-aware semantic acceptance avoids rejecting semantically valid tokens due to token-level mismatch, thereby increasing the accepted length per round and reducing device–edge interactions, as illustrated in Fig. 2(b). By explicitly incorporating CSI, the acceptance policy adapts to channel variability and achieves a more efficient communication–computation trade-off.
\begin{algorithm}[t]
\caption{WISV decoding with dual hidden-upload protocols under greedy verification}
\label{alg:WISV}
\begin{algorithmic}[1]
\REQUIRE Draft model $D$, target model $T$ at the edge, decision head $g_\theta$, window size $K$, threshold $\tau$, protocol $Proto \in \{\mathsf{FH},\mathsf{SH}\}$
\ENSURE Output sequence $\mathbf{y}$

\STATE $\mathbf{y} \leftarrow [\,]$
\WHILE{generation not finished}
    \STATE Obtain draft tokens and hidden states: $(\tilde{\mathbf{y}}, \mathbf{H}^D) \leftarrow Draft(D,\mathbf{y},K)$
    
    \IF{$Proto=\mathsf{FH}$}
        \STATE $Upload(\tilde{\mathbf{y}}, \mathbf{H}^D)$
    \ELSE
        \STATE $Upload(\tilde{\mathbf{y}})$
    \ENDIF

    \STATE Obtain target logits and hidden states: $(\mathbf{L}, \mathbf{H}^T) \leftarrow Verify(T,\mathbf{y},\tilde{\mathbf{y}})$
    \STATE Get candidate mismatch positions: $\mathcal{S} \leftarrow Localize(\mathbf{L}, \tilde{\mathbf{y}})$
    
    \IF{$Proto=\mathsf{SH}$}
        \STATE $RequestHidden(\mathcal{S})$
        \STATE $\{\mathbf{h}^D_i\}_{i\in\mathcal{S}} \leftarrow Upload(\mathbf{H}^D,\mathcal{S})$
    \ENDIF
    
    \STATE $\hat{j}\leftarrow \min\{\, i\in\mathcal{S}:\ \sigma(g_\theta([\mathbf{h}^D_i;\mathbf{h}^T_i;\mathbf{c}])) \ge \tau \,\}$ \COMMENT{or $\hat{j}=\emptyset$ if none}

    \IF{$\hat{j}=\emptyset$}
        \STATE $y_{\mathrm{next}} \leftarrow \arg\max \mathbf{L}_{K+1}$
        \STATE Full accept: commit $\tilde{\mathbf{y}}_{1:K}$ and the next token
        \STATE $\mathbf{y}\leftarrow \mathbf{y}\ \Vert\ \tilde{\mathbf{y}}_{1:K}\ \Vert\ [y_{\mathrm{next}}]$
    \ELSE
        \STATE Commit prefix and correct: $y^\star_{\hat{j}}\leftarrow \arg\max \mathbf{L}_{\hat{j}}$
        \STATE Downlink feedback: send $\langle \hat{j}, y^\star_{\hat{j}}\rangle$
        \STATE $\mathbf{y}\leftarrow \mathbf{y}\ \Vert\ \tilde{\mathbf{y}}_{1:\hat{j}-1}\ \Vert\ [y^\star_{\hat{j}}]$
    \ENDIF
\ENDWHILE
\RETURN $\mathbf{y}$
\end{algorithmic}
\end{algorithm}

\begin{figure*}[t]
\centering
\subfloat[]{
  \includegraphics[width=0.62\textwidth]{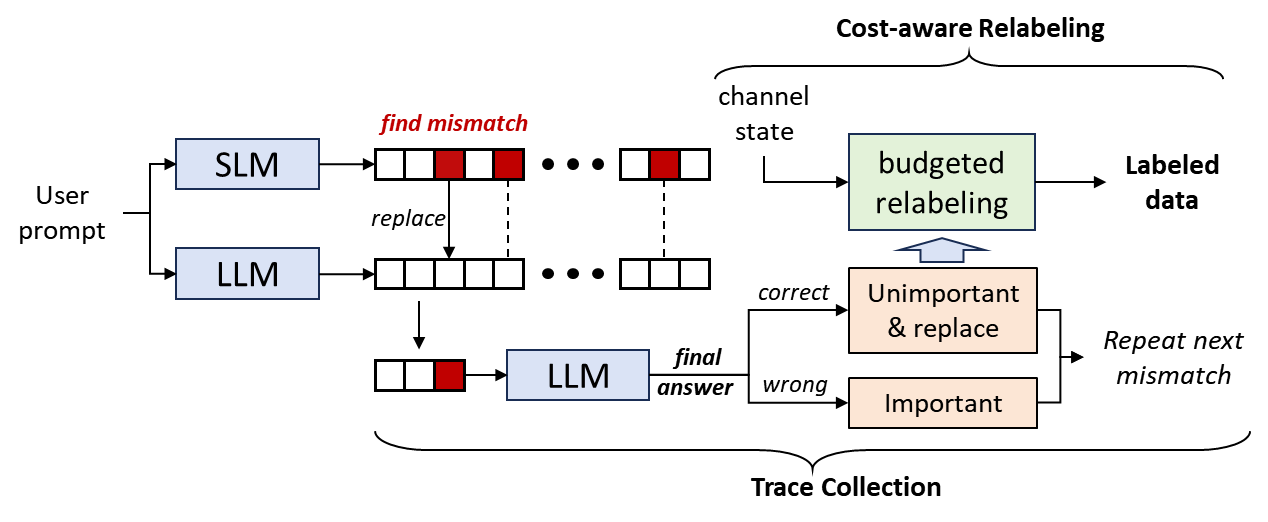}
  \label{fig:pipeline}
}\hspace{0.01\textwidth}
\subfloat[]{
  \includegraphics[width=0.33\textwidth]{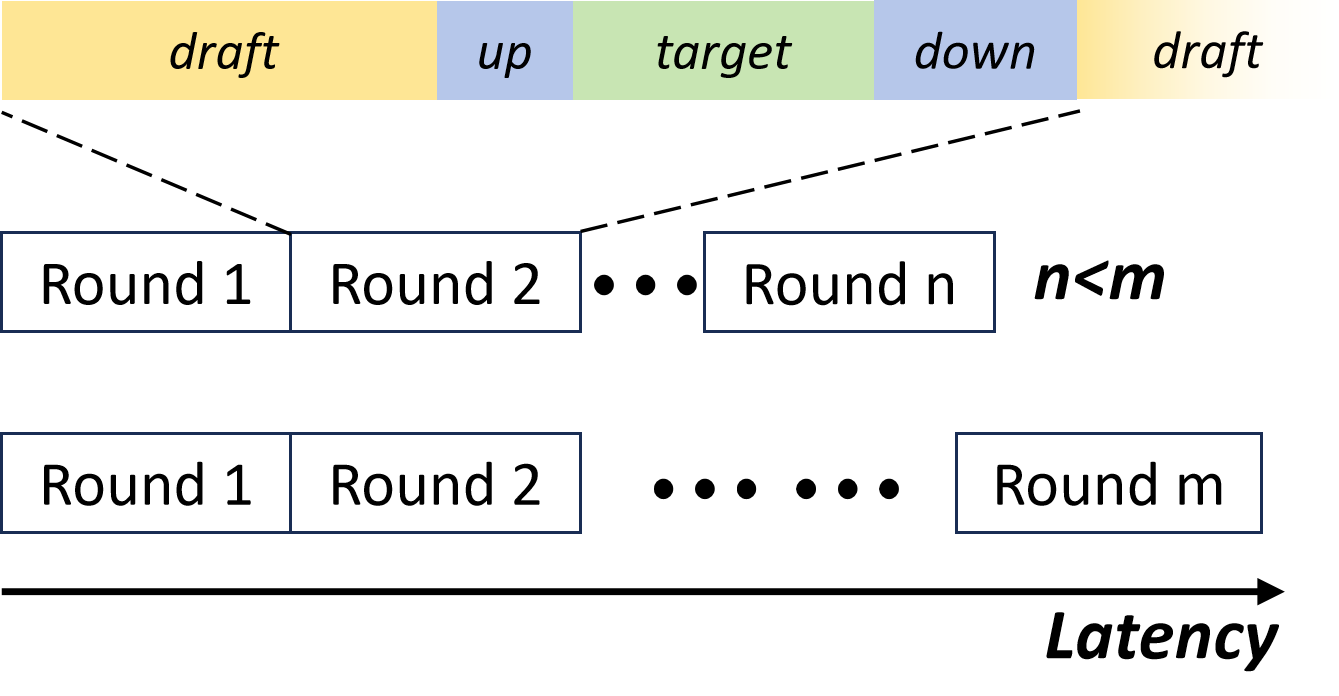}
  \label{fig:latency}
}
\caption{(a) The wireless-informed supervised dataset construction pipeline for training the WISV head. (b) End-to-end latency decomposition across interaction rounds.}
\label{fig:xxx}
\end{figure*}

\section{Problem formulation and performance metrics}
\label{sec:problem}
In this section, we formulate the communication and computation models for device–edge speculative decoding under wireless constraints. We first characterize the communication cost and latency under different hidden-state transmission protocols. We then model the computation cost of the drafter and target models. Finally, we introduce performance metrics and present an optimization-oriented view that captures the trade-off between semantic acceptance and communication overhead.
\subsection{Communication Model}
\label{sec:comm_model}
We model the device-edge link as a bidirectional wireless channel, whose instantaneous condition at round $r$ is characterized by CSI $\mathbf{c}^{r}$. In particular, the CSI captures both the effective link rate and link reliability, including the achievable uplink/downlink transmission rates and the packet error rate (PER).
Let $R_{\mathrm{u}}(\mathbf{c}^{r})$ and $R_{\mathrm{d}}(\mathbf{c}^{r})$ denote the achievable uplink and downlink rates, respectively, and let $\mathrm{PER}_{\mathrm{u}}(\mathbf{c}^{r})$ and $\mathrm{PER}_{\mathrm{d}}(\mathbf{c}^{r})$ denote the corresponding uplink and downlink packet error rates.

The system operates in interaction rounds indexed by $r=1,2,\ldots$, where the drafter proposes a token block of length $K_r\le K$, and the edge returns feedback at the first rejected position.

Let $|\mathcal{V}|$ be the vocabulary size and $b_{\mathrm{id}}=\lceil \log_2 |\mathcal{V}| \rceil$ be the number of bits required to represent a token ID. Let $d_h$ be the hidden-state dimension of the drafter model and $b_h$ the number of bits per hidden element (e.g., $b_h=16$ for FP16), so that the cost of uploading one drafter hidden vector is
\begin{equation}
B_{\mathrm{hid}} = d_h \, b_h \quad \text{(bits per token hidden)}.
\end{equation}

We also include fixed protocol overheads $B_{\mathrm{hdr,u}}$ and $B_{\mathrm{hdr,d}}$ for uplink and downlink messages.

At each round, the edge returns the first rejected position index $\hat{j}_r$ and the corrected token ID $y_{\hat{j}_r}^\star$. 
Let $b_{\mathrm{pos}}$ be the bitwidth for encoding a position index. The feedback payload is
\begin{equation}
B_{\mathrm{fb}} = B_{\mathrm{hdr,d}} + b_{\mathrm{pos}} + b_{\mathrm{id}}
\label{eq:B_fb}
\end{equation}

In the full-hidden (FH) option, the drafter uploads the $K_r$ token IDs and \emph{all} drafter hidden states in a single uplink message. The uplink payload is
\begin{align}
B_{\mathrm{u}}^{\mathrm{FH}}(r) 
&= B_{\mathrm{hdr,u}} 
+ K_r \, b_{\mathrm{id}}
+ K_r \, B_{\mathrm{hid}}
\label{eq:B_u_FH}
\end{align}

We decompose communication latency into payload-dependent transmission terms and a round-based interaction overhead. The former captures the expected serialization time under the effective link rate and packet errors, while the latter, denoted by $\tau_{\mathrm{RTT}}$, captures round-level non-payload delay, including propagation, bidirectional handshake, queueing, and protocol/network-stack processing overhead. The communication latency is

\begin{equation}
\begin{aligned}
T_{\mathrm{comm}}^{\mathrm{FH}}(r)
=&\ \frac{B_{\mathrm{u}}^{\mathrm{FH}}(r)}
{R_{\mathrm{u}}(\mathbf{c}^{r})\left(1-\mathrm{PER}_{\mathrm{u}}(\mathbf{c}^{r})\right)} \\
&+ \frac{B_{\mathrm{fb}}}
{R_{\mathrm{d}}(\mathbf{c}^{r})\left(1-\mathrm{PER}_{\mathrm{d}}(\mathbf{c}^{r})\right)}
+ \tau_{\mathrm{RTT}} .
\end{aligned}
\label{eq:T_comm_FH}
\end{equation}

In the selective-hidden (SH) option, the drafter first uploads token IDs. After the edge performs a forward pass and mismatch localization, it requests drafter hidden states only for a subset of positions.
Let $m_r \in \{0,1,\ldots,K_r\}$ denote the number of requested positions in round $r$, and let the downlink request include these indices.
The first uplink payload is
\begin{equation}
B_{\mathrm{u1}}^{\mathrm{SH}}(r)
= B_{\mathrm{hdr,u}} + K_r\, b_{\mathrm{id}}
\label{eq:B_u1_SH}
\end{equation}
The downlink request payload is
\begin{equation}
B_{\mathrm{req}}(r)
= B_{\mathrm{hdr,d}} + m_r \, b_{\mathrm{pos}}
\label{eq:B_req}
\end{equation}
The second uplink payload (requested hidden states) is
\begin{equation}
B_{\mathrm{u2}}^{\mathrm{SH}}(r)
= B_{\mathrm{hdr,u}} + m_r \, B_{\mathrm{hid}}
\label{eq:B_u2_SH}
\end{equation}
Thus, the total communication latency under SH is

\begin{equation}
\begin{aligned}
T_{\mathrm{comm}}^{\mathrm{SH}}(r)
=&\ \frac{B_{\mathrm{u1}}^{\mathrm{SH}}(r)}
{R_{\mathrm{u}}(\mathbf{c}^{r})\left(1-\mathrm{PER}_{\mathrm{u}}(\mathbf{c}^{r})\right)} \\
&+ \frac{B_{\mathrm{req}}(r)}
{R_{\mathrm{d}}(\mathbf{c}^{r})\left(1-\mathrm{PER}_{\mathrm{d}}(\mathbf{c}^{r})\right)} \\
&+ \frac{B_{\mathrm{u2}}^{\mathrm{SH}}(r)}
{R_{\mathrm{u}}(\mathbf{c}^{r})\left(1-\mathrm{PER}_{\mathrm{u}}(\mathbf{c}^{r})\right)} \\
&+ \frac{B_{\mathrm{fb}}}
{R_{\mathrm{d}}(\mathbf{c}^{r})\left(1-\mathrm{PER}_{\mathrm{d}}(\mathbf{c}^{r})\right)}
+ 2\tau_{\mathrm{RTT}} .
\end{aligned}
\label{eq:T_comm_SH}
\end{equation}

Compared with FH, SH reduces uplink payload from $K_r B_{\mathrm{hid}}$ to $m_r B_{\mathrm{hid}}$ at the cost of an additional request/response exchange, i.e., potentially one additional RTT per round.

\subsection{Computation Model}
\label{sec:comp_model}
We model the computation latency of the drafter on UE and the target model on the edge server using a FLOPs-based abstraction. Let $P_D$ and $P_T$ denote the peak compute throughput (FLOPs/s) of the UE and the edge server, respectively. 
To account for non-ideal hardware utilization due to memory bandwidth bottlenecks, kernel launch overhead, KV-cache access, and software runtime effects, we use effective utilization factors $\alpha_D,\alpha_T\in(0,1]$. Consequently, the effective throughput is $\alpha_D P_D$ and $\alpha_T P_T$. For a model component requiring $F$ FLOPs, the execution time is modeled as
\begin{equation}
T_{\mathrm{comp}}(F;P,\alpha) = \frac{F}{\alpha P}.
\label{eq:Tcomp_general}
\end{equation}

Let $L_r$ be the length of the committed prefix at the start of round $r$, and $K_r$ be the length of the draft block generated in that round. 
The accepted length before the first rejection is denoted by $A_r \in [0,K_r]$, such that $L_{r+1} = L_r + A_r + 1$, assuming a correction token is appended at the first rejected position.

The drafter generates $K_r$ tokens autoregressively  utilizing KV caching. We denote $f_D(\ell)$ as the FLOPs required to generate a single new token given a current context length $\ell$. The total drafter FLOPs in round $r$ is
\begin{equation}
F_D(r) = \sum_{i=0}^{K_r-1} f_D(L_r+i).
\label{eq:FD_round}
\end{equation}

Following standard approximations for decoder-only Transformers, $f_D(\ell)$ is decomposed into the costs of attention, MLP, and output-head. For a model with $N_D$ layers, hidden dimension $d_D$ and feed-forward dimension $d^{\mathrm{ff}}_D$, a high-fidelity proxy for these costs is
\begin{equation}
f_D(\ell) \approx N_D \Big( c_1 d_D^2 + c_2 d_D d_D^{\mathrm{ff}} + c_3 \ell d_D \Big) + c_4 d_D |\mathcal{V}|,
\label{eq:f_token_proxy}
\end{equation}
where $c_1,c_2,c_3,c_4$ are implementation-specific constants that account for factors such as projection packing and fused kernels. The $\ell d_D$ term captures the incremental attention cost associated with KV-cache access, while the $d_D^2$ and $d_D d_D^{\mathrm{ff}}$ terms represent the linear projections and feed-forward operations.

The drafter compute time in round $r$ is
\begin{equation}
T_D(r) = \frac{F_D(r)}{\alpha_D P_D}.
\label{eq:TD_round}
\end{equation}

On the edge server, the target model validates the $K_r$ draft tokens, typically via a single parallelized forward pass conditioned on the prefix $L_r$. Let $f_T(\ell)$ denote the target FLOPs for processing one token at context length $\ell$, defined analogously to \eqref{eq:f_token_proxy} using target model dimensions $(N_T,d_T,d_T^{\mathrm{ff}})$.
Because the $K_r$ tokens form a causal chain, the target verification FLOPs is
\begin{equation}
F_T(r) = \sum_{i=0}^{K_r-1} f_T(L_r+i),
\label{eq:FT_round}
\end{equation}
and the corresponding compute time is
\begin{equation}
T_T(r) = \frac{F_T(r)}{\alpha_T P_T}.
\label{eq:TT_round}
\end{equation}
The target forward pass also computes the decoding-head logits for mismatch localization and correction; these overheads are subsumed within $f_T(\cdot)$.

The proposed decision head is designed to be significantly more lightweight than the standard linear head. Let $m_r$ be the number of positions flagged for evaluation. For a two-layer MLP with hidden width $d_J$, the FLOPs per position is on the order of
\begin{equation}
f_J \approx 2d_{\mathrm{in}} d_J + d_J + 2d_J \cdot 1 + 1,
\end{equation}
where $d_{\mathrm{in}}$ represents the dimension of the input feature vector comprising CSI and hidden states. The total decision overhead is $F_J(r) = m_r f_J$, with an execution time $T_J(r)=F_J(r)/(\alpha_T P_T)$. While typically negligible compared to $T_T(r)$, this term is included for high-precision latency modeling.

Combining communication and computation, the per-round latency for FH and SH is modeled as
\begin{align}
T^{\mathrm{FH}}(r) &= T_D(r) + T_{\mathrm{comm}}^{\mathrm{FH}}(r) + T_T(r) + T_J(r), \\
T^{\mathrm{SH}}(r) &= T_D(r) + T_{\mathrm{comm}}^{\mathrm{SH}}(r) + T_T(r) + T_J(r).
\end{align}
The overall end-to-end latency is given by $T_{\mathrm{E2E}}=\sum_{r=1}^{N_{\mathrm{round}}} T^{(\cdot)}(r)$, where $N_{\mathrm{round}}$ denotes the total number of interaction rounds required for sequence termination.

\subsection{Metrics}
\label{sec:metrics_objective}

We evaluate the proposed device-edge co-inference system from the perspectives of both task quality and system efficiency. In addition to conventional metrics such as accuracy and throughput, we introduce execution-specific indicators, including the average accepted length (AAL), round count, and end-to-end inference latency.

Let $D=\{(\mathbf{x}^{(n)},a^{(n)})\}_{n=1}^N$ denote an evaluation dataset containing $N$ prompts, where $x^{(n)}$ represents the $n$-th prompt and $a^{(n)}$ is its corresponding ground-truth answer or evaluation criterion. For each prompt $n$, the system generates an output sequence $\hat{\mathbf{y}}^{(n)}$ across $N_{\mathrm{round}}^{(n)}$ interaction rounds. 
In any given round $r$, the drafter proposes a block of length $K_r^{(n)}$, and the system accepts $A_r^{(n)}\in[0,K_r^{(n)}]$ tokens prior to the first rejection or rollback event.

While the global acceptance rate measures the overall fraction of accepted tokens, the AAL characterizes the average length of the accepted token span per speculation round. This directly reflects the efficiency of each verification cycle:
\begin{equation}
\mathrm{AAL}^{(n)} = \frac{1}{N_{\mathrm{round}}^{(n)}} \sum_{r=1}^{N_{\mathrm{round}}^{(n)}} A_r^{(n)},
\qquad
\mathrm{AAL} = \frac{1}{N}\sum_{n=1}^N \mathrm{AAL}^{(n)}.
\label{eq:aal}
\end{equation}

The interaction round count quantifies the frequency of device-edge exchanges required to complete a response:
\begin{equation}
\overline{N}_{\mathrm{round}} = \frac{1}{N}\sum_{n=1}^N N_{\mathrm{round}}^{(n)}.
\label{eq:round_count}
\end{equation}
Since each interaction round inherently introduces at least one RTT and associated serialization delays, $\overline{N}_{\mathrm{round}}$ serves as a critical system performance indicator under wireless channel conditions and closely correlates with the actual end-to-end latency.

Let $T_{\mathrm{E2E}}^{(n)}$ denote the total end-to-end wall-clock time required to generate $\hat{\mathbf{y}}^{(n)}$, factoring in device-side drafting, wireless transmissions, edge-side verification, and feedback exchanges. 
Building upon the per-round latency model derived in Section~\ref{sec:comp_model}, we express the latency as
\begin{equation}
T_{\mathrm{E2E}}^{(n)} = \sum_{r=1}^{N_{\mathrm{round}}^{(n)}} T^{(\cdot)}(r;n),
\qquad
\overline{T}_{\mathrm{E2E}} = \frac{1}{N}\sum_{n=1}^N T_{\mathrm{E2E}}^{(n)}.
\label{eq:e2e_time}
\end{equation}

\section{CSI-Aware semantic verification}
\label{sec:semantic}
This section details a lightweight, CSI-aware decision head and its trace-driven training pipeline. By dynamically balancing semantic risk against communication costs, it minimizes device-edge interactions under varying wireless conditions.
\subsection{Decision Head Design}
To transcend strict token-level alignment while maintaining negligible edge-side overhead, we introduce a lightweight \emph{CSI-aware decision head} integrated into the target model's backbone. 
This head performs token-level semantic rejection prediction: by processing the hidden states of a candidate draft token from both drafter and target models alongside the current CSI, it outputs a scalar rejection probability $p_i$ indicating how likely the token should be rejected and repaired rather than committed.

For each queried position $i$ in round $r$, we construct a feature vector
\begin{equation}
\mathbf{z}_i^{(r)} = \big[\, \mathbf{h}^{D}_i \ ;\ \mathbf{h}^{T}_i \ ;\ \mathbf{c}^{r} \,\big] \in \mathbb{R}^{d_{\text{in}}},
\label{eq:feature_concat}
\end{equation}
where $\mathbf{c}^{r}$ represents the CSI feature vector. Unless otherwise specified, $\mathbf{h}^{D}_i$ and $\mathbf{h}^{T}_i$ are extracted from the final-layer hidden states of the drafter and target Transformer backbones at token position $i$, respectively.

The decision head is implemented as a two-layer MLP with a small hidden width, followed by a sigmoid function:
\begin{align}
s_i^{(r)} &= g_{\theta}\!\left(\mathbf{z}_i^{(r)}\right), \qquad 
p_i^{(r)} = \sigma\!\left(s_i^{(r)}\right),
\label{eq:head_forward}
\end{align}
where $g_{\theta}$ denotes the MLP, $s_i^{(r)}\in\mathbb{R}$ is the logit, and $p_i^{(r)}\in(0,1)$ is the rejection probability.
Specifically, $g_{\theta}$ consists of a linear projection, ReLU activation, dropout regularization, and a final linear layer:
\begin{equation}
g_{\theta}(\mathbf{z}) 
= \mathbf{w}_2^\top\, \mathrm{Dropout}\!\left(\mathrm{ReLU}(\mathbf{W}_1 \mathbf{z} + \mathbf{b}_1)\right) + b_2,
\label{eq:mlp_form}
\end{equation}
with $\mathbf{W}_1\in\mathbb{R}^{d_J\times d_{\text{in}}}$ and $\mathbf{w}_2\in\mathbb{R}^{d_J}$. Given the small hidden dimension $d_J$, this head introduces only $\mathcal{O}(d_{\text{in}}d_J)$ parameters and negligible FLOPs compared to the target model's forward pass, ensuring scalability for high-throughput edge serving.

The decision head is trained offline via supervised learning on speculative decoding traces. Each training example consists of a feature-label pair $(\mathbf{z}, y)$, where $\mathbf{z}$ is the feature vector and label $y\in\{0,1\}$ indicates whether the token should be rejected under the target policy. $y=1$ means that the draft token should be rejected, and $y=0$ means that it can be accepted under the desired policy.
We minimize the standard binary cross-entropy (BCE) loss:
\begin{equation}
\mathcal{L}_{\text{BCE}}(\theta)
= - \mathbb{E}\big[\, y \log p + (1-y)\log(1-p) \,\big],
\label{eq:bce}
\end{equation}
incorporating weight decay and class balancing where necessary. This offline approach decouples head optimization from expensive end-to-end execution, facilitating efficient iteration over feature sets and thresholds.

During runtime, the head evaluates candidate tokens and generates $p_i^{(r)}$. We apply a decision threshold $\tau \in (0,1)$ to obtain the binary rejection decision:
\begin{equation}
\hat{a}_i^{(r)} = \mathbb{I}\left(p_i^{(r)} \ge \tau\right).
\label{eq:threshold}
\end{equation}
The first position where $\hat{a}_i^{(r)}=1$ triggers a rollback and subsequent feedback as described in Section~\ref{sec:arch_workflow}. 
The threshold $\tau$ serves as a tunable parameter for the latency-quality trade-off: a lower $\tau$ enforces conservative (fewer risky acceptances) validation, while a higher $\tau$ prioritizes communication efficiency by increasing the accepted span, which is particularly advantageous under adverse CSI conditions.

\subsection{Supervision / Labeling}
The training dataset is synthesized from speculative decoding traces. The primary challenge lies in the fact that correction decisions are not purely governed by distribution alignment; rather, they must balance the risk of semantic degradation against the CSI-dependent cost of additional interaction rounds. We address this via a trace-driven pipeline followed by wireless-informed cost-aware relabeling as shown in Fig.~\ref{fig:pipeline}.

\textbf{Stage 1: Trace collection:}
We execute speculative decoding on a calibration corpus, following a methodology inspired by AutoJudge \cite{AJ}. For each prompt, we record an \emph{episode} consisting of the ordered mismatch positions $\mathcal{P}=\{p_1<p_2<\cdots<p_T\}$ encountered when comparing draft and target tokens under the same prefix, together with the corresponding hidden representations required by the decision head.

For each mismatch position $p_t$, we store: $(i)$ the drafter-side hidden state $\mathbf{h}^D_{p_t}$, $(ii)$ the target hidden state $\mathbf{h}^T_{p_t}$, and $(iii)$ mismatch metadata, including the position index and the target/draft token IDs. Unless otherwise specified, hidden states are taken from the final Transformer layer.

We further assign a task-driven base label to each mismatch, indicating whether it is likely to affect downstream quality. Following prior work on judge-based decoding, this is obtained by replacing draft/target tokens at mismatch positions and evaluating whether the final task outcome changes.
The resulting binary label $b_t\in\{0,1\}$ indicates whether mismatch $t$ is critical, where a ``critical'' mismatch refers to one whose acceptance would alter the downstream result.

\textbf{Stage 2: Wireless-informed cost-aware relabeling.}
Given an episode $\mathcal{P}$ with mismatch positions $\{p_t\}_{t=1}^{T}$ and a sampled CSI state $\mathbf{c}$, we relabel each mismatch by a binary action $a_t\in\{0,1\}$, where $a_t=1$ denotes \emph{repair} and $a_t=0$ denotes \emph{skip}.
The relabeling is made \emph{wireless-informed}: under favorable CSI, original importance labels are preserved, while under constrained channels, controlled relaxation is introduced to reflect communication-computation limitations.

\paragraph{Smoothed importance}
Let $b_t\in\{0,1\}$ be the original importance indicator. To capture the contextual influence of critical mismatches, we define a smoothed importance score
\begin{equation}
\tilde b_t \;=\; \max_{k:\,b_k=1}\alpha^{|t-k|},\quad \alpha\in(0,1).
\label{eq:smooth_importance_short}
\end{equation}
This formulation allows critical mismatches to influence nearby positions with exponential decay.
We also summarize channel quality by a scalar $q(\mathbf{c})\in[0,1]$.

\paragraph{CSI-aware budgeted relabeling objective}
We then introduce a CSI-dependent repair budget $B(\mathbf{c})$, which is monotone with respect to channel quality. The relabeling is formulated as a budget-constrained optimization problem:
\begin{equation}
\begin{aligned}
\min_{a_t} \quad & \sum_{t=1}^{T} (b_t-a_t)\,\tilde b_t \\
\text{s.t.}\quad
& \sum_{t=1}^{T} a_t \;\le\; B(\mathbf{c}),\\
& a_t \in \{0,1\},\quad a_t \le b_t,\ \forall t,
\end{aligned}
\label{eq:budget_opt_short}
\end{equation}
where the constraint $a_t\le b_t$ enforces one-way relabeling, i.e., only originally important mismatches can be relaxed.

\paragraph{Randomized Lagrangian policy}
To enable efficient large-scale relabeling, we adopt a Lagrangian relaxation with multiplier $\lambda(\mathbf{c})$:
\begin{equation}
\mathcal{L}(a_t;\lambda)
=
\sum_{t=1}^{T} (b_t-a_t)\tilde b_t
+
\lambda\!\left(\sum_{t=1}^{T} a_t - B(\mathbf{c})\right).
\label{eq:lagrangian_short}
\end{equation}
This yields a separable decision rule with threshold structure:
$
a_t^\star
=
b_t\,\mathbf{1}\!\left\{\tilde b_t > \lambda(\mathbf{c})\right\}.
$

To further obtain a smooth and sampleable policy, we introduce an entropy-regularized Bernoulli relaxation:
$
\pi_t
\;=\;
\mathbb{P}(a_t=1 \mid b_t,\mathbf{c}).
\label{eq:pi_def}
$
We optimize the expected relaxed objective with an entropy regularizer. This yields the closed-form policy
\begin{equation}
\pi_t
=
b_t \cdot
\sigma\!\left(
\frac{\tilde b_t - \lambda(\mathbf{c})}{\rho}
\right),
\label{eq:soft_policy}
\end{equation}
where $\sigma(\cdot)$ is the logistic function and $\rho>0$ controls the smoothness.
As channel conditions deteriorate, $\lambda(\mathbf{c})$ increases, reducing repair probability and yielding adaptive label relaxation.

\paragraph{Training instances}
We sample multiple CSI realizations per episode and apply the CSI-conditioned randomized policy \eqref{eq:soft_policy} to obtain relabeled actions $\{a_t\}$.
Each mismatch yields a supervised instance $(\mathbf{x}_t,y_t)$ with $y_t=a_t$ and
\[
\mathbf{x}_t=[\mathbf{h}_{D,t};\mathbf{h}_{T,t};\mathbf{c}],
\]
where $\mathbf{h}_{D,t}$ and $\mathbf{h}_{T,t}$ denote the drafter and target hidden representations aligned to mismatch $t$, respectively.
This produces link-aware labels that are conservative under good channels and appropriately relaxed under constrained wireless conditions.

\subsection{Inference Integration}
To reduce unnecessary computation, we do not apply the decision head to all positions within a speculative block. Instead, we first perform mismatch localization to construct a candidate set $\mathcal{S}_r \subseteq \{t,\ldots,t+K_r-1\}$, which contains positions most likely to require rejection or repair.
In the greedy verification setting, $\mathcal{S}_r$ is obtained by comparing draft tokens with the target model’s argmax predictions and collecting mismatch indices. This filtering step significantly reduces the number of evaluated positions, with $m_r = |\mathcal{S}_r|$ typically satisfying  $m_r \ll K_r$.

The proposed integration is inherently modular. The decision head does not replace the target model’s decoding head; rather, it acts as an auxiliary verification module that intervenes only at critical positions where strict token-level matching would prematurely terminate the speculative block. This design ensures compatibility with existing speculative decoding frameworks and preserves the correctness of the base verification pipeline.

From a computational perspective, the additional overhead is minimal. Since the target forward pass is required regardless of whether WISV is enabled, the incremental cost arises solely from evaluating a lightweight decision head on the candidate set. Importantly, this small overhead leads to a disproportionate system-level gain. By avoiding unnecessary early rejections, WISV increases the accepted length per round and reduces the number of rollback-triggered interaction rounds. Under wireless constraints, where each additional round incurs nontrivial communication latency, this reduction directly translates into significant end-to-end latency improvement.

\section{Simulation Results}
\label{sec:sim}
In this section, we evaluate WISV under wireless device–edge settings using both task-level and system-level metrics. We present comparisons with speculative decoding baselines and ablation studies to demonstrate the benefits of CSI-aware semantic verification in improving efficiency and latency.

\subsection{Experiment Settings}
We evaluate the proposed WISV using \textsc{Llama-3.2-1B-Instruct} as the on-device drafter and \textsc{Llama-3.1-8B-Instruct} as the edge-side target model on the GSM8K benchmark \cite{gsm8k}. 
We adopt a link-layer abstraction rather than explicitly simulating physical-layer effects such as Rayleigh fading or AWGN. This abstraction is widely used in wireless edge system studies to focus on communication–computation coupling at the application layer. In particular, the device–edge link state at each interaction round is characterized by two parameters: the effective data rate $R$, capturing net throughput after protocol overhead and scheduling, and the packet error rate (PER), capturing link-level reliability.

\subsection{Performance Evaluation}
We compare WISV against representative speculative decoding baselines in terms of task accuracy, end-to-end latency, throughput, and related metrics defined in Section~\ref{sec:metrics_objective}.

\begin{figure}[t]
    \centering
    \includegraphics[width=1\linewidth]{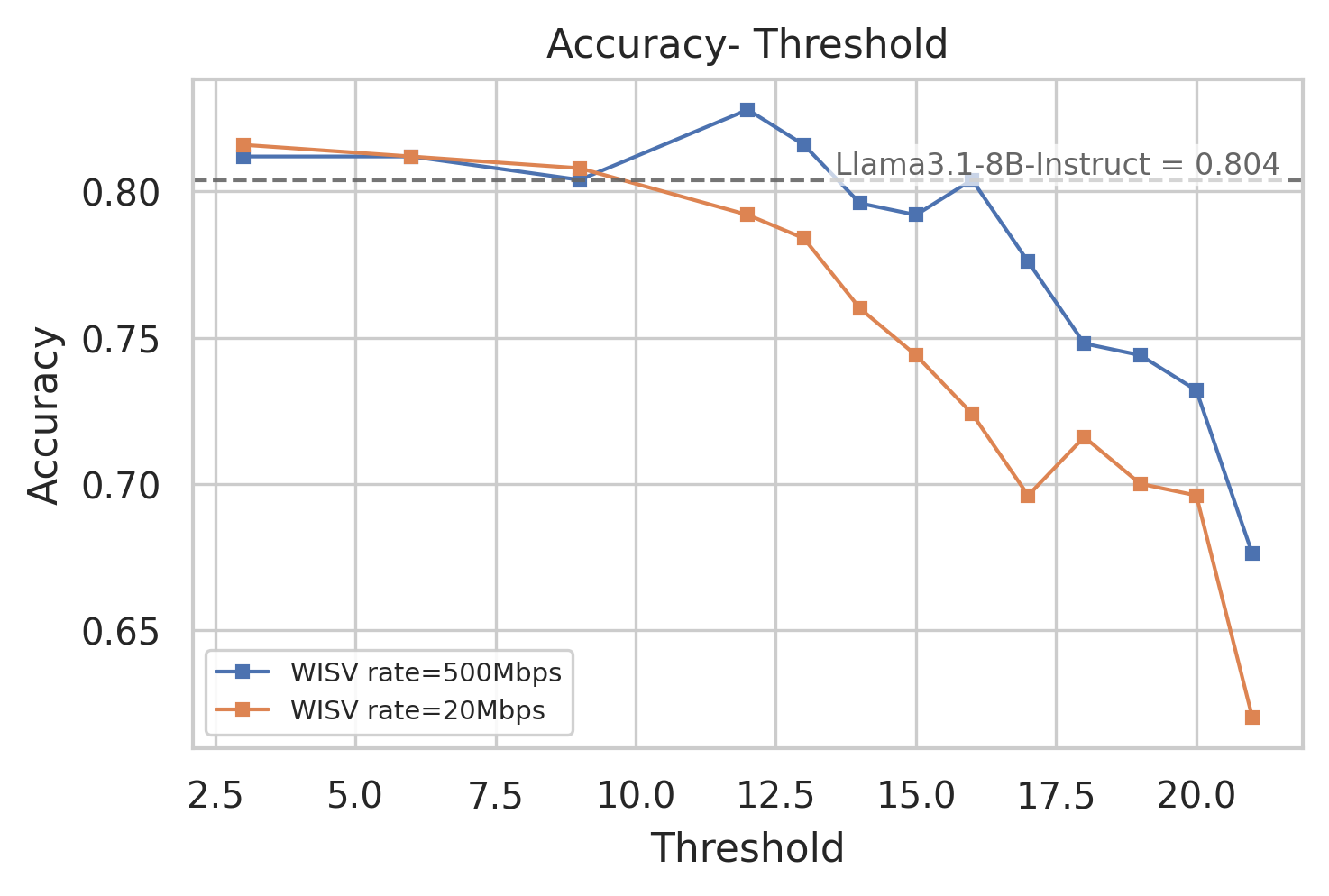}
    \caption{The accuracy of the proposed WISV under different thresholds.}
    \label{fig:acc_thr}
\end{figure}

\emph{1) Accuracy:} Fig.~\ref{fig:acc_thr} shows the GSM8K accuracy under different acceptance thresholds of the decision head. Since the threshold is applied to the rejection probability, a larger threshold makes rejection harder and thus yields a more permissive policy. The dashed horizontal line represents the accuracy of target-only inference (80.4\%), which serves as the reference. The threshold controls the permissiveness of the decision head: a larger value leads to more aggressive acceptance of draft tokens. When the threshold is 0, WISV reduces to standard speculative decoding, whereas at the maximum value, all draft tokens are accepted.

We observe that when the threshold is below 9, WISV maintains accuracy comparable to or slightly exceeding the target-only baseline across all channel conditions. As the threshold increases beyond this point, accuracy degradation gradually appears, particularly under poor channel conditions, due to more aggressive acceptance of semantically uncertain tokens. At threshold 14, the degradation under poor channels is approximately 5\%, while accuracy under favorable channels remains stable. When the threshold exceeds 15, accuracy drops sharply in all scenarios, indicating that overly permissive policies compromise generation reliability.

Based on this accuracy–latency tradeoff, we select threshold 12 for subsequent experiments. Under this setting, WISV achieves 79.6\% accuracy under poor channels (less than 1\% degradation) and 82.8\% under favorable channels, outperforming the target-only baseline.

\begin{figure*}[t]
\centering
\subfloat[500\,Mbps, RTT = 50\,ms]{%
  \includegraphics[width=0.49\textwidth]{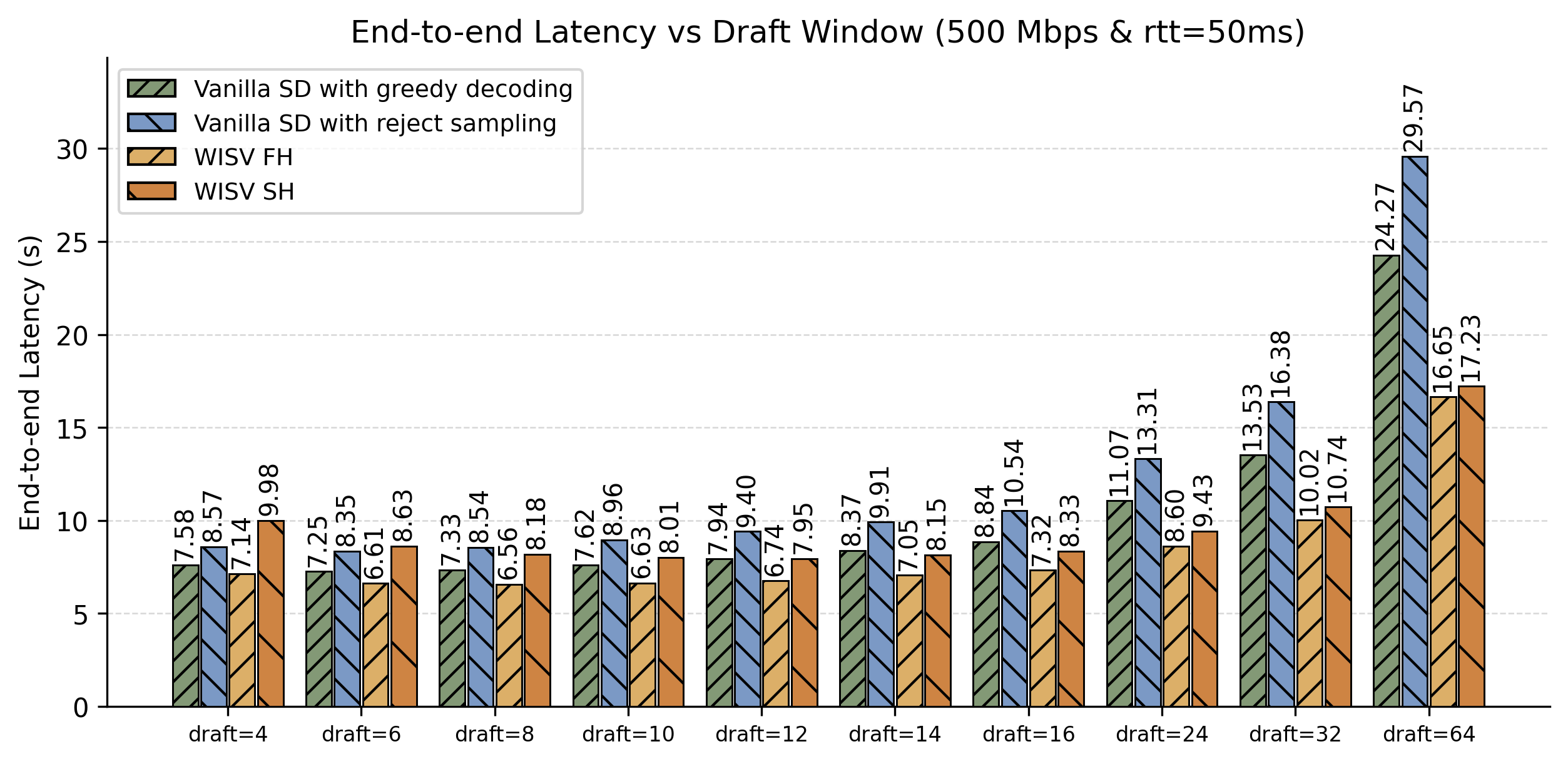}
  \label{fig:IL_bar_500_50}
}\hfill
\subfloat[20\,Mbps, RTT = 50\,ms]{%
  \includegraphics[width=0.49\textwidth]{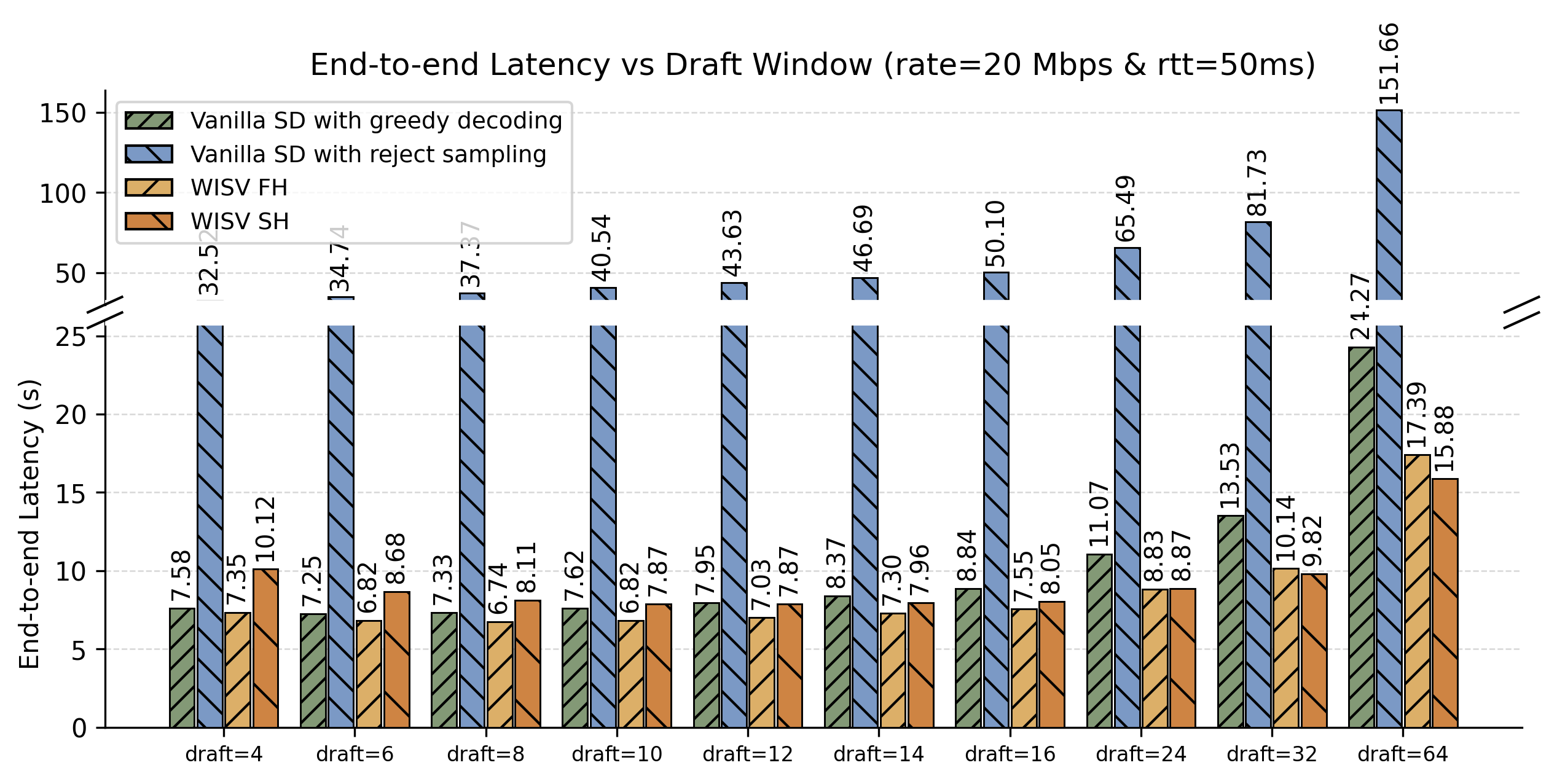}
  \label{fig:IL_bar_20_50}
}\\[4pt]
\subfloat[500\,Mbps, RTT = 5\,ms]{%
  \includegraphics[width=0.49\textwidth]{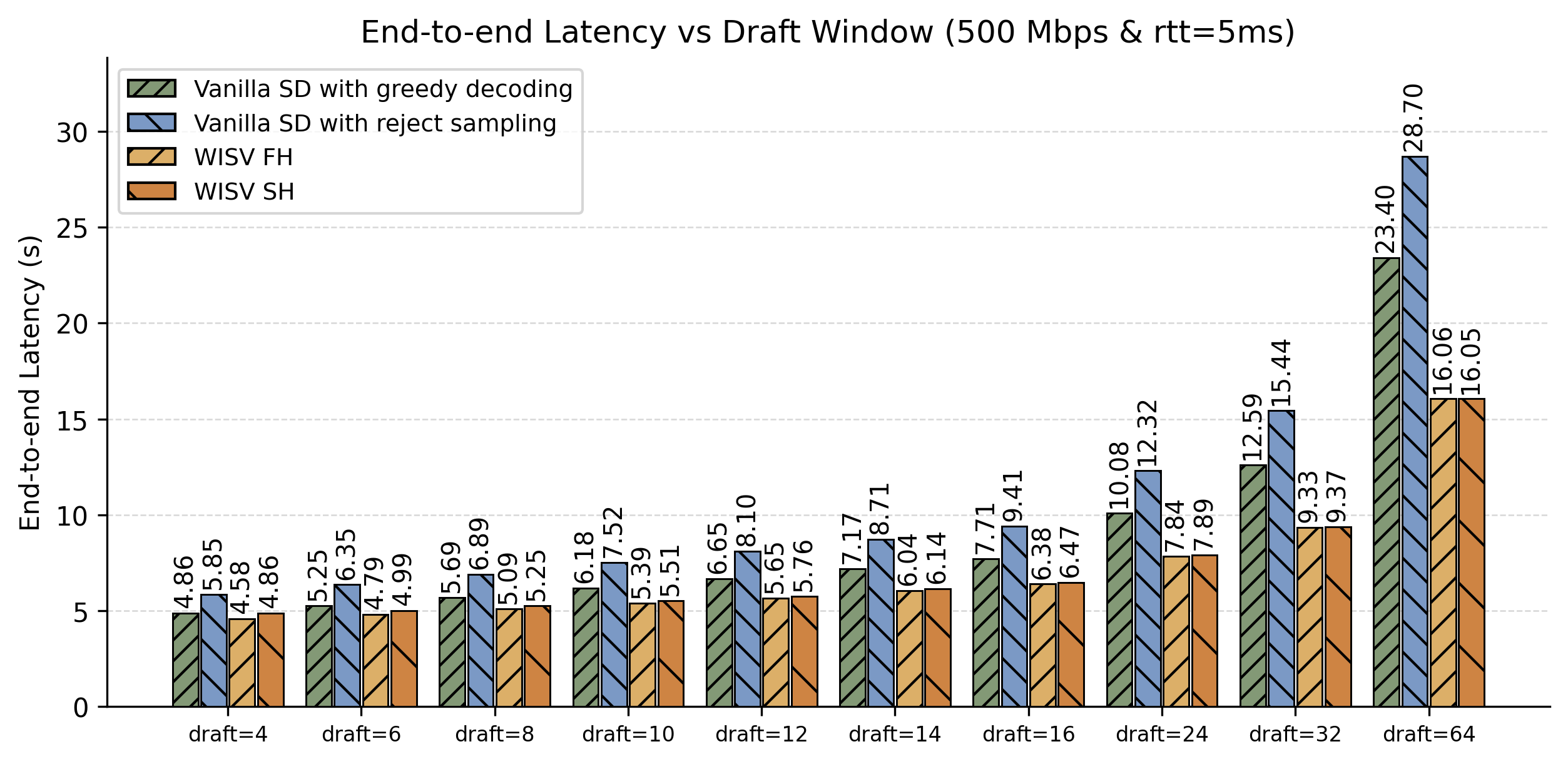}
  \label{fig:IL_bar_500_5}
}\hfill
\subfloat[20\,Mbps, RTT = 5\,ms]{%
  \includegraphics[width=0.49\textwidth]{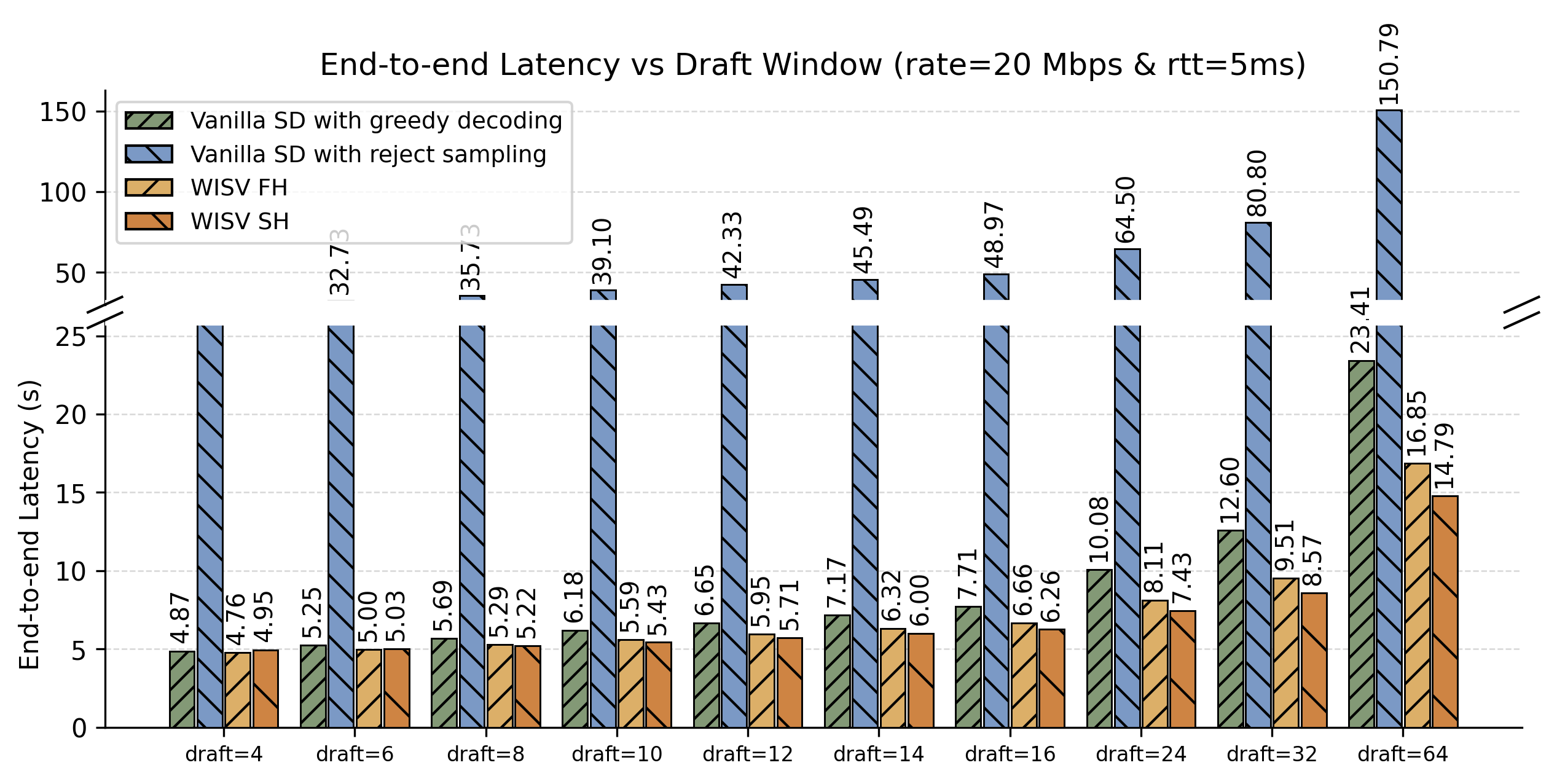}
  \label{fig:IL_bar_20_5}
}
\caption{Average end-to-end latency of different methods under different transmission rates and RTT settings.}
\label{fig:IL_bar}
\end{figure*}

\begin{table*}[t]
\centering
\caption{Comparison of different methods across two channel rates.}
\label{tab:two_scenarios}
\setlength{\tabcolsep}{3pt}
\renewcommand{\arraystretch}{1.15}

\begin{tabularx}{\textwidth}{l l *{8}{>{\centering\arraybackslash}X}}
\toprule
\multicolumn{2}{c}{\multirow{2}{*}{Method}}
& \multicolumn{4}{c}{Rate=500Mbps} & \multicolumn{4}{c}{Rate=20Mbps} \\
\cmidrule(lr){3-6}\cmidrule(lr){7-10}
\multicolumn{2}{c}{}
& Latency$\downarrow$ & AAL$\uparrow$ & Rounds$\downarrow$ & Throughput$\uparrow$
& Latency$\downarrow$ & AAL$\uparrow$ & Rounds$\downarrow$ & Throughput$\uparrow$ \\
\midrule

\multirow{3}{*}{draft=10}
& SD greedy  & 7.616 & 6.607 & 31.912 & 27.683 & 7.617 & 6.607 & 31.912 & 27.681 \\
& SD reject  & 8.962 & 6.595 & 32.036 & 23.577 & 40.542 & 6.586 & 32.040 & 5.205 \\
& WISV       & 6.634 & 7.478 & 27.736 & 31.267 & 6.816 & 7.805 & 27.144 & 31.083 \\
\midrule

\multirow{3}{*}{draft=16}
& SD greedy  & 8.842 & 8.336 & 25.128 & 23.689 & 8.843 & 8.336 & 25.128 & 23.687 \\
& SD reject  & 10.543 & 8.305 & 25.244 & 19.886 & 50.096 & 8.335 & 25.112 & 4.179 \\
& WISV       & 7.317 & 9.991 & 20.732 & 28.309 & 7.553 & 10.630 & 19.936 & 28.057 \\
\midrule

\multirow{3}{*}{draft=24}
& SD greedy  & 11.074 & 9.513 & 22.020 & 18.916 & 11.075 & 9.513 & 22.020 & 18.915 \\
& SD reject  & 13.313 & 9.486 & 22.132 & 15.770 & 65.493 & 9.505 & 22.072 & 3.203 \\
& WISV       & 8.604 & 12.142 & 17.044 & 24.052 & 8.832 & 13.230 & 15.976 & 23.930 \\
\midrule

\multirow{3}{*}{draft=32}
& SD greedy  & 13.525 & 10.128 & 20.680 & 15.484 & 13.527 & 10.128 & 20.680 & 15.483 \\
& SD reject  & 16.376 & 10.060 & 20.848 & 12.806 & 81.731 & 10.104 & 20.744 & 2.564 \\
& WISV       & 10.021 & 13.522 & 15.260 & 20.592 & 10.140 & 14.979 & 13.892 & 20.522 \\
\midrule

\multirow{3}{*}{draft=64}
& SD greedy  & 24.272 & 10.789 & 19.272 & 8.566 & 24.274 & 10.789 & 19.272 & 8.566 \\
& SD reject  & 29.570 & 10.708 & 19.424 & 7.033 & 151.663 & 10.763 & 19.368 & 1.374 \\
& WISV       & 16.653 & 15.489 & 13.152 & 12.233 & 17.394 & 17.343 & 12.092 & 12.057 \\
\bottomrule
\end{tabularx}
\end{table*}

\emph{2) End-to-End Latency:} Fig.~\ref{fig:IL_bar} compares end-to-end inference latency under different wireless transmission rates and RTT settings. We consider four representative scenarios: poor channels (20 Mbps) and good channels (500 Mbps), each with RTTs of 5 ms and 50 ms.

Each subplot presents four methods: greedy speculative decoding, rejection-sampling–based speculative decoding, and two WISV variants (SH and FH). Rejection sampling exhibits the highest latency due to full-vocabulary probability transmission, while greedy decoding incurs negligible communication overhead by transmitting only token IDs.

WISV introduces two protocol options with different communication–interaction tradeoffs. When RTT is large (50 ms), SH suffers from an additional interaction round, resulting in higher latency than FH and even slightly worse performance than greedy decoding. In contrast, under low-rate conditions (20 Mbps), FH becomes slightly slower than SH at large draft windows due to increased uplink payload from full hidden-state transmission. This reveals a clear regime-dependent preference: SH is more suitable for low-rate and low-RTT scenarios, whereas FH is preferable under high-rate or high-RTT conditions.

Overall, FH consistently achieves lower latency than greedy decoding across all settings. Moreover, latency exhibits a non-monotonic trend with respect to the draft window size: increasing the window initially reduces latency by decreasing interaction rounds, but further enlargement leads to diminishing returns due to limited gains in acceptance length. At 500Mbps, WISV FH achieves up to 31.4\% lower end-to-end latency compared with greedy speculative decoding.

\emph{3) Average Accepted Length, Round Count and Throughput:}
Table~\ref{tab:two_scenarios} shows the AAL, round count, and throughput under different wireless transmission rates with an RTT of 50 ms and threshold 12. Because SH and FH share the same verification logic, they exhibit identical AAL and round counts; thus, we present FH only. Compared with baseline methods, WISV consistently achieves a higher AAL across all draft window sizes, which directly translates into fewer interaction rounds. This improvement remains robust as the draft window increases, indicating that WISV effectively mitigates premature rejection.

Across channel conditions, the decision head adapts its acceptance policy based on CSI. Under poor channels (20 Mbps), it becomes more permissive, leading to further increases in AAL and corresponding reductions in round count. Although FH incurs higher uplink payload, its end-to-end latency remains only slightly higher than that under good channels, while still significantly outperforming both greedy and rejection-based speculative decoding.
At 20 Mbps, WISV FH achieves up to 60.8\% higher AAL, 37.3\% fewer interaction rounds compared with greedy speculative decoding.

\subsection{Ablation Study}
\begin{figure}[t]
    \centering
    \includegraphics[width=1\linewidth]{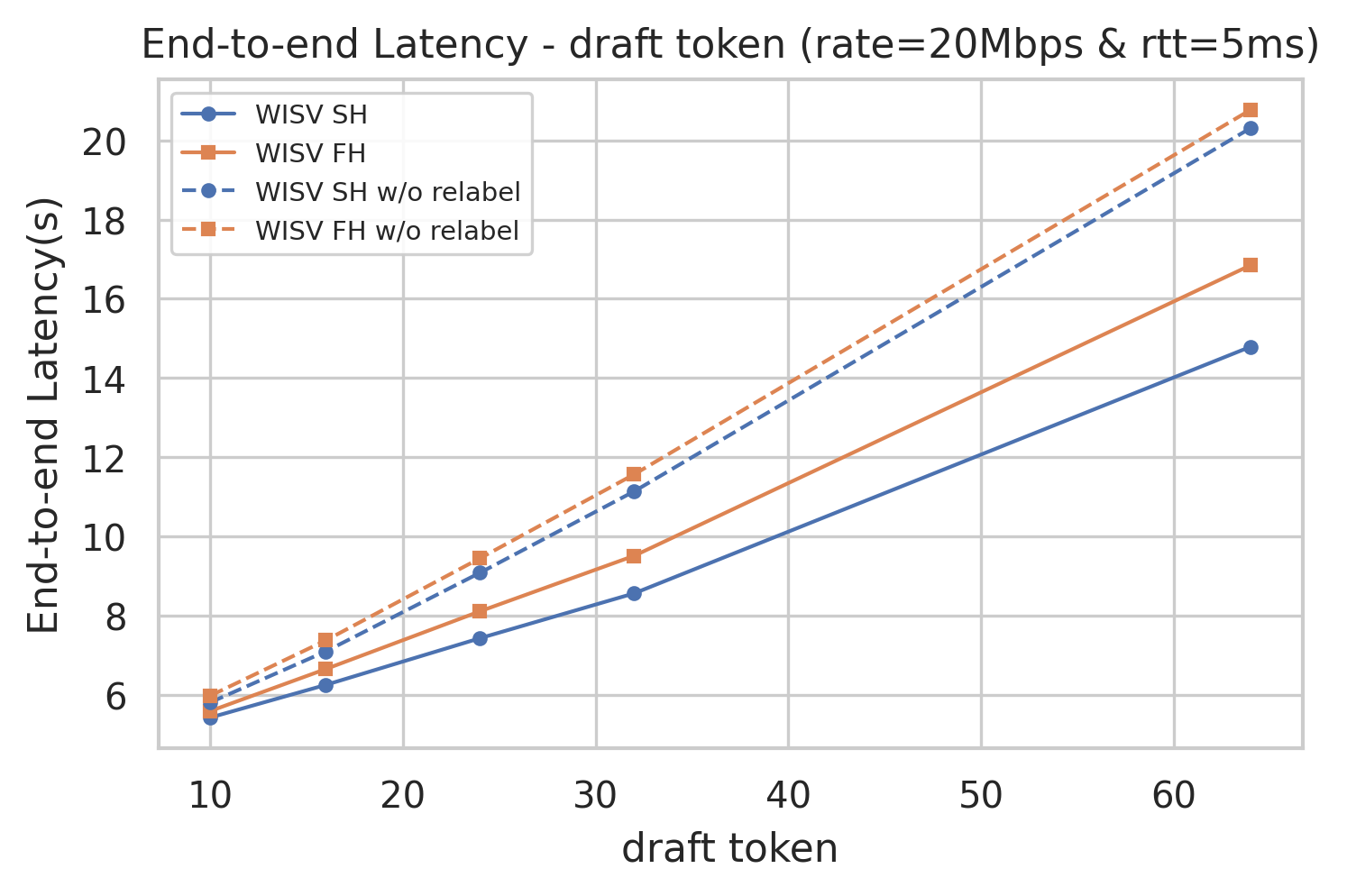}
    \caption{The ablation study of WISV. }
    \label{fig:ablation}
\end{figure}

Fig.~\ref{fig:ablation} presents the ablation results. The dashed curves correspond to the decision head trained without CSI-aware budgeted relabeling. Incorporating CSI enables the decision head to adapt its acceptance policy to channel conditions by adjusting its permissiveness. In particular, under poor channels, the model becomes more permissive, resulting in a higher average accepted length and consequently lower end-to-end latency.

\section{Hardware Testbed for WISV}
\label{sec:hardware}
To validate WISV in real-world settings, we build a hardware testbed for device–edge distributed inference and evaluate its performance under dynamic wireless conditions in this section.
\subsection{System Components}
Our hardware prototype follows a practical device–edge co-inference architecture for distributed speculative decoding. The lightweight drafter is deployed on an NVIDIA Jetson AGX Orin 64GB, while the target model runs on an edge server equipped with an NVIDIA A40 GPU. The two sides communicate over a Wi-Fi link via a TCP-based client–server protocol. For each decoding request, the device performs local drafting and transmits the required information to the edge, which then executes target-side verification and returns feedback. This setup reflects a realistic wireless inference pipeline and enables direct measurement of communication–computation coupling in distributed speculative decoding.

\begin{figure}
    \centering
    \includegraphics[width=1\linewidth]{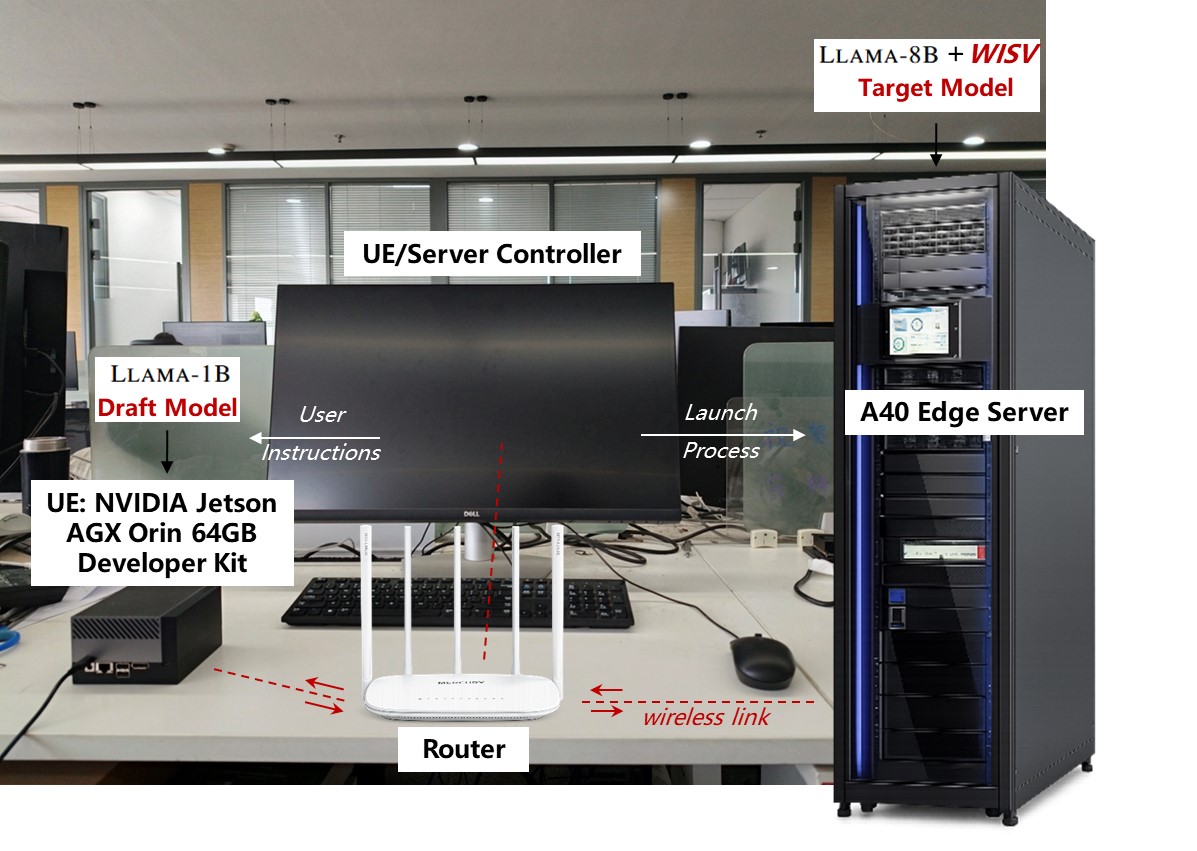}
    \caption{System components of the hardware testbed.}
    \label{fig:hardware}
\end{figure}

Fig.~\ref{fig:hardware} illustrates the hardware testbed. The Jetson device acts as the UE running \textsc{Llama-3.2-1B-Instruct}, while the edge server hosts \textsc{Llama-3.1-8B-Instruct} for verification. As a further supplement, we also adopt the Qwen series of models to demonstrate the effectiveness and generality of the proposed method. Specifically, we deploy \textsc{Qwen2.5-0.5B-Instruct} as the draft model on the end device and \textsc{Qwen2.5-7B-Instruct} as the target model on the edge server. 

\subsection{Hardware Implementation}
The hardware implementation differs from the simulation in two aspects. First, instead of fixing the protocol (FH or SH), the system adaptively selects the protocol based on real-time channel conditions. Second, the hardware testbed operates under dynamic wireless conditions rather than preset rates and RTT.

Specifically, before each request, the UE measures the current RTT via a lightweight handshake and selects the protocol accordingly: FH is used when RTT exceeds 10 ms, while SH is used otherwise. This design captures the tradeoff between payload size and interaction rounds—FH reduces round trips at the cost of higher uplink payload, whereas SH is more efficient under low-latency links. In addition, to introduce greater uncertainty in the wireless channel and to emulate the mobility characteristics of end devices in realistic scenarios, we place the UE at different distances from the router during the experiments. Specifically, when the distance between the UE and the router is 20 meters, the channel quality is relatively poor, and the measured average uplink rate is 172 Mbps. When the distance is 2 meters, the channel quality is better, with a measured average uplink rate of 498 Mbps.

\begin{figure*}[t]
\centering
\subfloat[]{%
  \includegraphics[width=0.49\textwidth]{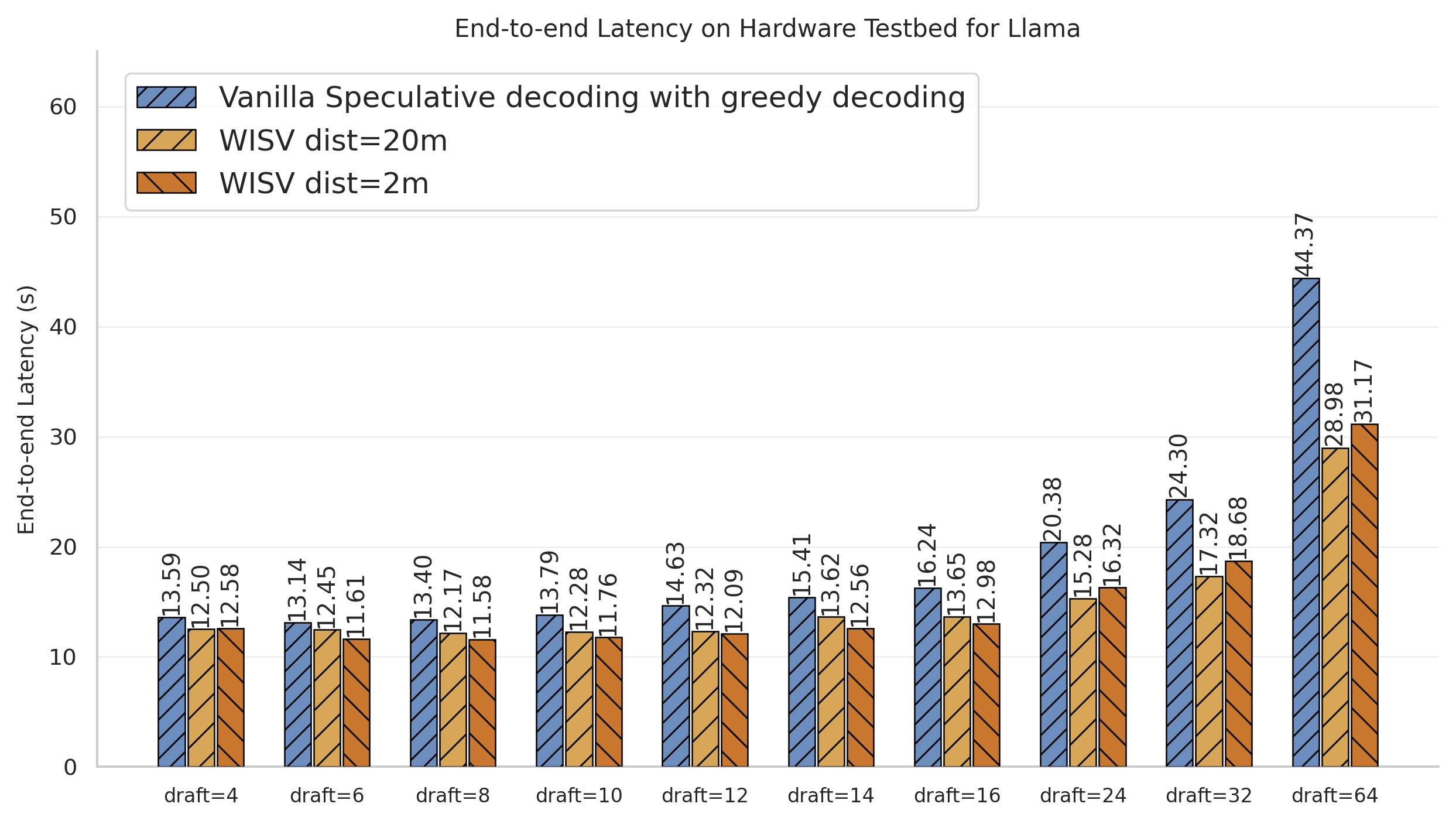}
  \label{fig:jetson_latency_llama}
}\hfill
\subfloat[]{%
  \includegraphics[width=0.49\textwidth]{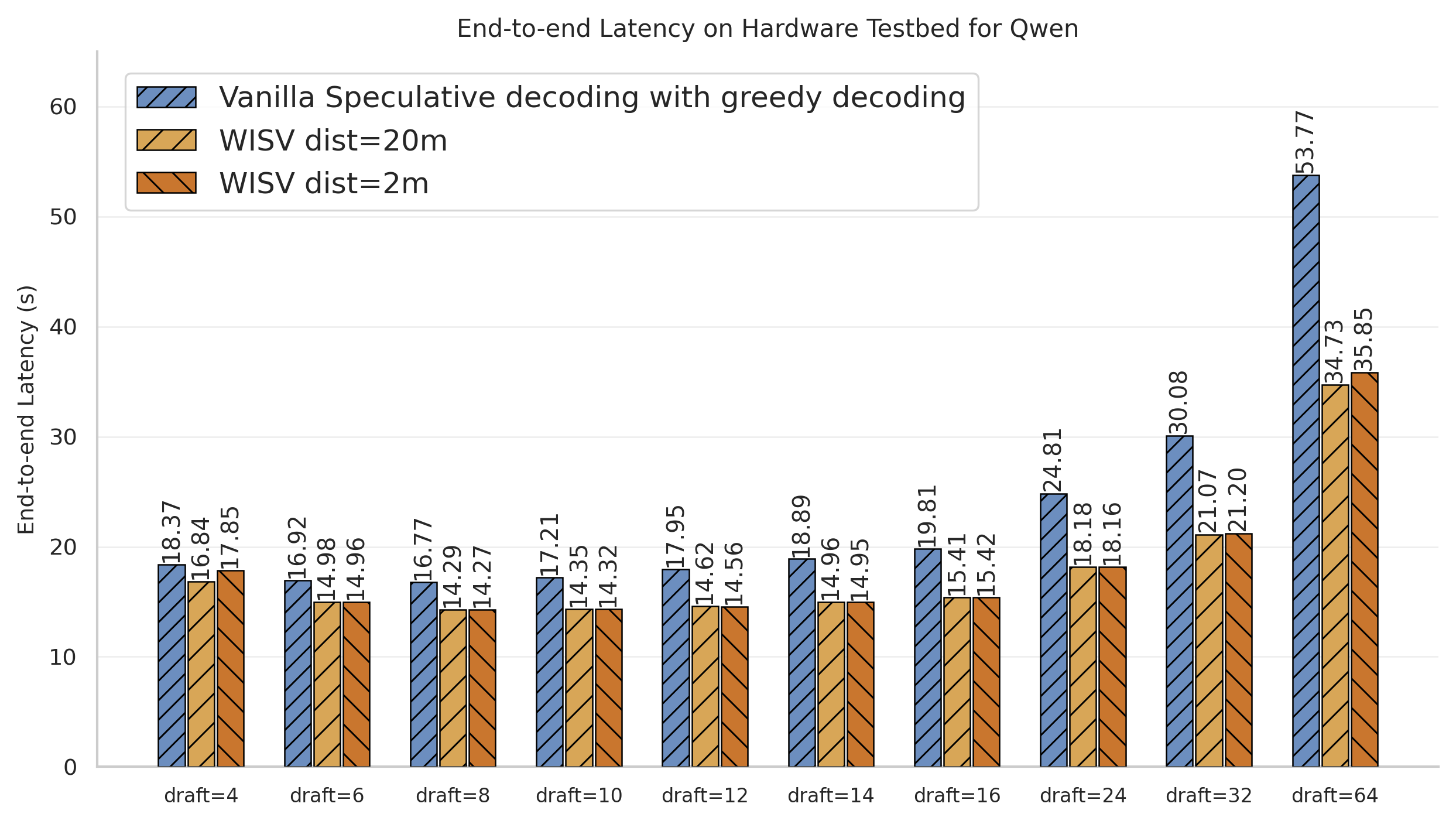}
  \label{fig:jetson_latency_qwen}
}
\caption{End-to-end latency on the hardware testbed with adaptive protocol selection for different model settings.}
\label{fig:jetson_latency}
\end{figure*}

Fig.~\ref{fig:jetson_latency} presents the end-to-end latency measured on the hardware testbed with adaptive protocol selection. All results are obtained by tuning the threshold of the WISV decision head to ensure inference accuracy. For the Llama series, the accuracy remains no lower than 79.6\% across all draft window sizes, corresponding to a drop of less than 1\% compared with the 80.4\% accuracy of standard speculative decoding, which matches target-only accuracy on GSM8K. For the Qwen series, the accuracy remains no lower than 86\% across all draft window sizes, corresponding to an accuracy drop of about 1.3\% compared with the 87.3\% accuracy of standard speculative decoding. 
The left subfigure corresponds to the Llama series, while the right subfigure corresponds to the Qwen series. In each group, the left bar represents standard greedy speculative decoding in the distributed scenario. The middle bar corresponds to the case where the distance between the UE and the router is 20 meters, while the right bar corresponds to the case where the distance is 2 meters. Compared with standard speculative decoding, WISV consistently achieves lower latency across all draft window sizes. This improvement holds for both small and large draft windows, demonstrating the robustness of the proposed wireless-informed verification mechanism in real deployment conditions.

As the draft window increases, latency exhibits a U-shaped trend for both methods, reflecting the tradeoff between reduced interaction rounds and increased per-round computation and communication overhead. Notably, WISV maintains a clear advantage over the baseline throughout this range, and the gap becomes more pronounced at larger draft sizes, where reducing unnecessary rollbacks is more critical. A comparison of WISV under different UE-to-router distances shows that, with its channel-aware adaptive strategy selection, WISV can still achieve an end-to-end latency close to that under high-quality channel conditions even when the UE is placed 20 meters away from the router, corresponding to a poorer channel. In fact, at very small and very large draft window sizes, it can even slightly outperform the high-quality-channel case. This is because the WISV decision head tends to adopt a more permissive verification behavior under poorer channel conditions by taking channel state information as an input, i.e., it is more likely to accept the current draft token. Under poorer channel conditions, it relaxes verification to reduce the number of communication rounds, which not only lowers the communication component of latency, but also decreases the number of drafting rounds and target-model verification steps. Specifically, the maximum latency reduction reaches 35.41\% when the draft window is 64 under Qwen settings compared with standard greedy speculative decoding.
These results validate that the proposed wireless-informed semantic verification not only improves performance in simulation but also translates effectively to real-world device–edge systems under dynamic wireless conditions.

\section{Conclusion}
\label{sec:conclusion}
This paper presented WISV, a novel wireless-informed semantic verification framework designed to address the over-rejection challenge in distributed speculative decoding. By incorporating CSI-aware semantic acceptance, WISV significantly enhances token acceptance rates and minimizes interaction rounds, effectively reducing end-to-end latency without compromising generation quality. The lightweight nature of the proposed decision head ensures minimal computational overhead and seamless integration into existing LLM inference pipelines. Beyond individual performance gains, WISV enables the use of larger speculative windows, which alleviates edge server workloads and enhances system capacity for large-scale user requests. Our findings underscore the potential of exploiting communication-computation synergy to scale high-performance edge LLM inference services across heterogeneous wireless networks.

\bibliographystyle{IEEEtran}
\bibliography{refs}  

\end{document}